\documentclass[
    twocolumn,       
    superscriptaddress,
    amsmath,amssymb,
    aps,
    physrev
]{revtex4-2}

\usepackage{graphicx}
\usepackage{dcolumn}
\usepackage{bm}

\begin{document}

\title{
Coarse-Grained Dynamics with Spatial Disorder and Non-Markovian Memory
}

\author{Chuyi Liu}
\affiliation{School of Physics, Zhejiang University, Hangzhou 310058, China}
\affiliation{Shanghai Artificial Intelligence Laboratory, Shanghai 200232, China}

\author{Yifeng Guan}
\affiliation{School of Physics, Zhejiang University, Hangzhou 310058, China}
\affiliation{Shanghai Artificial Intelligence Laboratory, Shanghai 200232, China}

\author{Jingyuan Li}
\email{jingyuanli@zju.edu.cn}
\affiliation{School of Physics, Zhejiang University, Hangzhou 310058, China}
\affiliation{Institute for Advanced Study in Physics, Zhejiang University, Hangzhou 310027, China}

\author{Mao Su}
\email{sumao@pjlab.org.cn}
\affiliation{Shanghai Artificial Intelligence Laboratory, Shanghai 200232, China}
\affiliation{Shenzhen Institute of Advanced Technology, Chinese Academy of Sciences, Shenzhen 518055, China}

\date{\today}

\begin{abstract}
We introduce the spatial disorder-generalized Langevin equation (SD-GLE), a data-driven method for constructing coarse-grained (CG) dynamics in heterogeneous systems. Unlike conventional CG approaches that rely on a mean-field potential, SD-GLE utilizes a variational Bayesian framework with a random field prior to explicitly disentangle static spatial disorder from viscoelastic friction. Numerical results demonstrate the limits of standard GLEs, whereas SD-GLE accurately extrapolates long-time dynamics to capture the anomalous diffusion crossover from short trajectories and recover the ensemble statistical properties inherent to the disordered nature of these systems.

\end{abstract}

\maketitle

Accurately predicting the long-term dynamics of particles in complex heterogeneous systems---such as living cells \cite{2004_Bio_Anomalous,2021_Bio_Anomalous, 2012_living_cell}, cross-linked polymer networks \cite{2023_Polymer_Anomalous,2024_Polymer_Anomalous, 2012_polymer}, and glass-forming liquids \cite{2023_Glass_Anomalous, 2020_Glass_Anomalous}---remains a central challenge in non-equilibrium statistical mechanics. In these systems, anomalous transport typically emerges from the coupling of two distinct physical mechanisms \cite{2011_Two_Mechanisms,2010_Two_Mechanisms,2014_Diffusion_Random_Potential}: the viscoelastic response of the environment, which results in a non-Markovian friction, and static spatial disorder, which creates a rugged potential leading to localized trapping. Traditional phenomenological models, such as fractional Brownian motion (FBM) or continuous-time random walks (CTRW), generally assume a single underlying mechanism and thus struggle to quantitatively reproduce the observed dynamics \cite{2014_anomalous_diffusion,2015_two_mechanisms_toolbox}.

Data-driven methods have been introduced to accurately reconstruct the coupled dynamics directly from observations. 
Although existing studies have addressed the distinction between these two mechanisms of anomalous diffusion from observations \cite{2013_Indistinguishable, 2021_ML_Distinguish, 2023_Distinguish, 2021_diffusion_objective, 2019_classification,2019_measurement}, these efforts have primarily focused on a qualitative identification of their physical origins. 
Recent data-driven frameworks based on the generalized Langevin equation (GLE) \cite{Tepper2024MemoryKernel,2016_Data_Driven_GLE,2017_Memory,2022Likelihood_Memory,2023_HuanLei_Memory,2024_AIGLE,2024_coarse_grain, 2022_position_dependent,2021_data_driven_GLE} offer a systematic approach, enabling the joint inference of the effective potential energy surface and the viscoelastic memory kernel. However, in highly heterogeneous environments, these approaches encounter a fundamental limitation. Rugged potential energy surface induces long-lived trapping and weak ergodicity breaking \cite{2008_Ergodicity_Breaking,2011_Ergodicity_Breaking}, such that individual trajectories probe only local environments, leading to substantial inter-trajectory dispersion over finite-time ensembles \cite{2016_Single_trajectory}. Standard GLE methods rely on a global mean-field approximation of the effective potential, which averages out local spatial disorder. As a result, unresolved spatial heterogeneity is absorbed into the inferred memory effect, leading to an overestimation of temporal correlations and compromising the extrapolation of long-time transport properties \cite{2025_GLE_with_Potential}.
Thus, developing a framework that quantitatively disentangles the respective contributions of spatial disorder and temporal memory from limited trajectory data remains an unresolved problem. 

In this letter, we introduce the spatial disorder-generalized Langevin equation (SD-GLE) framework, which employs a non-parametric variational Bayesian inference approach to explicitly decouple static spatial correlations from viscoelastic memory. The heterogeneous potential energy surface is modeled as a Gaussian random field. By maximizing the marginal likelihood over all trajectories, SD-GLE leverages single-trajectory variational posteriors to adaptively capture local spatial fluctuations, while simultaneously extracting the viscoelastic friction kernel and the spatial correlation function of the potential on a global scale. Through validation in both ideal systems and simulated glass-forming liquids, we demonstrate that the proposed model accurately reproduces the disordered characteristics of the system. Furthermore, by incorporating the underlying spatial heterogeneity, simulations based on this framework correctly extrapolate long-time anomalous diffusion and naturally recover the essential features of disordered transport, including transient non-Gaussian displacement distributions and the massive inter-trajectory dispersion induced by weak ergodicity breaking.

We consider the dynamics of a low-dimensional coarse-grained (CG) coordinate \(Q(t)\) with conjugate momentum \(P(t)\), resulting from the projection of a many-body system onto a reduced subspace. Within the Mori-Zwanzig formalism \cite{2010_MZ_for_MD}, the effective dynamics takes the form:
\begin{equation}
\begin{aligned}
&\dot{Q}(t) = M^{-1}P(t),\\[4pt]
&\dot{P}(t) = -\nabla U(Q(t)) - \int_0^t K(t-s)V(s)\,ds + R(t).
\end{aligned}
\end{equation}
where $M$ is the mass matrix, $V = M^{-1}P$ is the velocity, and $K(t)$ and $R(t)$ denote the viscoelastic memory kernel and the fluctuating force respectively.

To model the spatial disorder, we employ a non-parametric Bayesian approach, treating the local effective potential $U(Q)$ as a realization of a Gaussian random field prior. This choice is motivated by the central limit theorem \cite{1990_CLT} for unresolved degrees of freedom, which universally governs the emergence of a Gaussian energy disorder in systems ranging from supercooled liquids \cite{1987_Supercool_GRF,2008_Supercool_GRF} to regulatory proteins on DNA tracks \cite{2014_Diffusion_Random_Potential,2016_Protein_GRF,2002_Protein_GRF,2004_Protein_GRF,2009_Protein_GRF,2012_Protein_GRF}. Specifically, $U(Q) \sim \mathcal {GP}(U_\mathrm{mean}(Q),C(Q,Q'))$, where $U_\mathrm{mean}$ refers to the mean field and $C(Q,Q')$ refers to the spatial correlation function.

We consider a discrete trajectory $X_{0:N}^{(\alpha)} = \{Q_n^{(\alpha)}, P_n^{(\alpha)} \}_{n=0}^N$ for time step $\Delta t$ and total time $T=N\Delta t$, where $\alpha$ indexes a specific realization of the local potential. The likelihood of this trajectory is then obtained by marginalizing over all possible configurations of $U^{(\alpha)}$

\begin{equation}
    p(X_{0:N}^{(\alpha)}; K,C)=\int \mathcal{D}[U^{(\alpha)}]p(X_{0:N}^{(\alpha)}\mid U^{(\alpha)};K)p(U^{(\alpha)};C).
\end{equation}

As this high-dimensional path integral is computationally intractable, we employ variational inference to maximize a lower bound on the log-likelihood (the evidence lower bound, ELBO). By introducing an approximate posterior distribution $q(U^{(\alpha)})$, the log-likelihood is strictly bounded by:

\begin{equation}
\begin{aligned}
\ln p(X_{0:N}^{(\alpha)};K,C) \ge& \mathbb{E}_{q(U^{(\alpha)})}[ \ln p(X_{0:N}^{(\alpha)} \mid U^{(\alpha)};K) ] \\&- D_{\text{KL}}(q(U^{(\alpha)}) \parallel p(U^{(\alpha)};C)).
\end{aligned}
\end{equation}

The direct parameterization of $q(U^{(\alpha)})$ is still unfeasible due to the continuous nature of the function space. However, evaluating the trajectory likelihood only requires the potential values at the discrete spatial coordinates visited in $X_{0:N}^{(\alpha)}$. To construct a tractable finite-dimensional representation, we introduce a set of $M\ll N$ learnable inducing variables $\mathbf{u}^{(\alpha)}$ to serve as a low-dimensional backbone for the local landscape \cite{2009_SVGP}. By defining the joint variational distribution as $q(U^{(\alpha)}, \mathbf{u}^{(\alpha)}) = p(U^{(\alpha)} \mid \mathbf{u}^{(\alpha)})q(\mathbf{u}^{(\alpha)})$, where $q(\mathbf{u}^{(\alpha)})=\mathcal{N}(\bm{\mu},\mathbf{S})$ is a Gaussian variational distribution, the marginal distribution over the potential field at the trajectory data locations analytically resolves to:

\begin{equation}
\begin{aligned}
q(U^{(\alpha)}) &= \int q(\mathbf{u}^{(\alpha)})\,p(U^{(\alpha)}|\mathbf{u}^{(\alpha)})\,d\mathbf{u}^{(\alpha)} \\
&= \mathcal{N}\!\left(U^{(\alpha)} \mid \mathbf{W}\bm{\mu},\; \mathbf{\Sigma}_{nn}+\mathbf{W}(\mathbf{S}-\mathbf{\Sigma}_{mm})\mathbf{W}^{\mathsf{T}}\right),
\end{aligned}
\end{equation}
where $\mathbf{W} = \mathbf{\Sigma}_{nm}\mathbf{\Sigma}_{mm}^{-1}$ acts as a weight matrix. Here, the subscripts $n$ and $m$ index the trajectory data coordinates and the inducing point locations, such that $\mathbf{\Sigma}_{nn}$, $\mathbf{\Sigma}_{mm}$, and $\mathbf{\Sigma}_{nm}$ denote the covariance matrices evaluated among these point sets (see Supplemental Material for details). Consequently, the intractable KL divergence in the ELBO strictly reduces to $D_{\text{KL}}(q(\mathbf{u}^{(\alpha)}) \parallel p(\mathbf{u}^{(\alpha)};C))$, yielding a fully differentiable objective function.

Beyond the spatial disorder, computing the exact trajectory probability $p(X_{0:N}^{(\alpha)}\mid U^{(\alpha)};K)$ also requires integrating out the viscoelastic memory. To this end, we embed the non-Markovian friction into an extended phase space by introducing auxiliary variables \(\bm{\zeta}\), which represent the degrees of freedom of the heat bath coupled to the CG coordinates \cite{2010_Markov_embedding,2010_Markov_Embedding_Temp}. The resulting Markovian process \([Q,P,\bm{\zeta}]\) is given by: 

\begin{equation}
\begin{aligned}
\dot{Q} &= P, \\
\dot{P} &= M^{-1}(-\nabla U(Q) + \mathbf{a}_p^{\mathsf T}\bm{\zeta}) \\
\dot{\bm{\zeta}} &= -\mathbf{A}\,\bm{\zeta} -\mathbf{a}_pV + \mathbf{B}\,\bm{\xi}(t),
\end{aligned}
\label{eq:extended}
\end{equation}
where \(\bm{\xi}(t)\) refers to the independent standard Gaussian noise. \(\{\mathbf{a}_p,\mathbf{A}\}\) parameterize the memory kernel \(K(t)=\mathbf{a}_p^{\mathsf T}e^{-\lvert t \rvert \mathbf{A}}\mathbf{a}_p\). For equilibrium processes, the fluctuation-dissipation theorem \(\mathbf{B}\mathbf{B}^{\mathsf T}=k_BT(\mathbf{A}+\mathbf{A}^{\mathsf T})\) should be satisfied \cite{2016_Data_Driven_GLE}. 

Conditioned on a specific local potential $U^{(\alpha)}$, evaluating the exact trajectory likelihood $p(X^{(\alpha)}_{0:N} \mid U^{(\alpha)};K)$ requires marginalizing over the unobservable auxiliary variables $\bm{\zeta}^{(\alpha)}_{0:N}$. Due to the conditionally linear and Markovian nature of Eq.~{\ref{eq:extended}}, this high-dimensional path integral can be evaluated exactly. Taking advantage of the explicit form of the transition probabilities, we apply an iterative predictor–corrector approach (also known as the Kalman filter)  \cite{1965_KF,2022Likelihood_Memory}. Starting from the trajectory up to step $k-1$, we determine the probability of the hidden variable $\bm{\zeta}^{(\alpha)}_k$ conditioned on the past information $X^{(\alpha)}_{0:k-1}$ with Eq.~{\ref{eq:extended}}. We then use the current observation $X^{(\alpha)}_k$ to correct this prediction, yielding the exact filtering distribution $p(\bm{\zeta}^{(\alpha)}_k \mid X^{(\alpha)}_{0:k},U^{(\alpha)};K)$. The filtering distribution provides the exact prior needed to evaluate the probability of the next trajectory step, $p(X^{(\alpha)}_{k+1} \mid X^{(\alpha)}_{0:k},U^{(\alpha)};K)$. As this recursive procedure runs forward from $k=0$ to $N$, it effectively orthogonalizes the temporal memory, projecting the correlated non-Markovian dynamics onto a sequence of independent white-noise innovations. Through the chain rule of probability, the intractable integration over the bath variables is bypassed, directly yielding the exact and fully differentiable marginal log-likelihood of the entire observed trajectory (see Supplemental Material for details).

Finally, integrating both the spatial and temporal marginalizations, the model parameters for spatial and temporal correlations $\theta_C,\theta_K$ and the posterior $q(\mathbf{u})$ are optimized by maximizing the ELBO:

\begin{equation}
\begin{aligned}
\mathcal{L}(\theta_C,\theta_K,q(\mathbf{u}))=&\sum_{\alpha}\mathbb{E}_{q(\mathbf{u}^{(\alpha)})}\mathbb{E}_{q(U\mid \mathbf{u}^{(\alpha)})}[\ln p(X_{0:N}^{(\alpha)}\mid U;\theta_K)]\\&-D_\text{KL}(q(\mathbf{u}^{(\alpha)})\|p(\mathbf{u}^{(\alpha)};\theta_C)).
\end{aligned}
\end{equation}

The expectation of the random potentials $U$ is evaluated using Monte Carlo method over the distribution of inducing points $u$, and the gradients are computed using the reparameterization trick. After optimization, the posterior distribution of the potential is used to generate an ensemble of potentials. Independent GLE trajectories are then simulated under these sampled potentials.

Let us first consider a one-dimensional system. To construct the disorder ensemble, we sample $64$ independent replicas of the potential energy surface from a Gaussian random field  characterized by the spatial correlation function $C(r - r') = \sigma^2 \exp\left( - \frac{|r - r'|^2}{2l^2} \right)$, where $\sigma$ sets the strength of the disorder, and $l$ is the correlation length. For each replica, we simulate 100 independent trajectories to thoroughly sample the local phase space. Each trajectory contains $4\times10^{3}$ steps with a time step of $5\times10^{-3}$.

We apply SD-GLE to the trajectory ensemble and compare it with a baseline standard GLE. To ensure a fair comparison, this baseline relies on a neural network for the global mean-field potential, but shares the identical auxiliary variable inference scheme for the memory kernel. As shown in Fig.~\ref{fig:1}(a), under the mean-field assumption, inference based on the standard GLE absorbs static spatial heterogeneity into the temporal dissipation term, leading to an overestimated memory kernel $K(t)$ with increased amplitude. By explicitly separating the static disorder from dynamic friction, SD-GLE accurately recovers the reference non-Markovian memory. This spatial resolution naturally reconstructs the rugged effective potential $U(r)$. Fig.~\ref{fig:1}(b) shows the posterior distribution of $U^{(\alpha)}$ inferred for a representative replica $\alpha$. The posterior mean faithfully tracks the reference landscape, explicitly resolving the local trapping environments. Regions of high uncertainty correctly reflect unexplored space where trajectory data is sparse. In contrast, lacking spatial decoupling, the standard GLE yields an unphysical, featureless potential. Furthermore, the spatial correlation function of the random field extracted via SD-GLE exactly matches the reference, confirming the thermodynamic consistency of the model across the ensemble.

\begin{figure}
\centering
\includegraphics[width=1.0\linewidth]{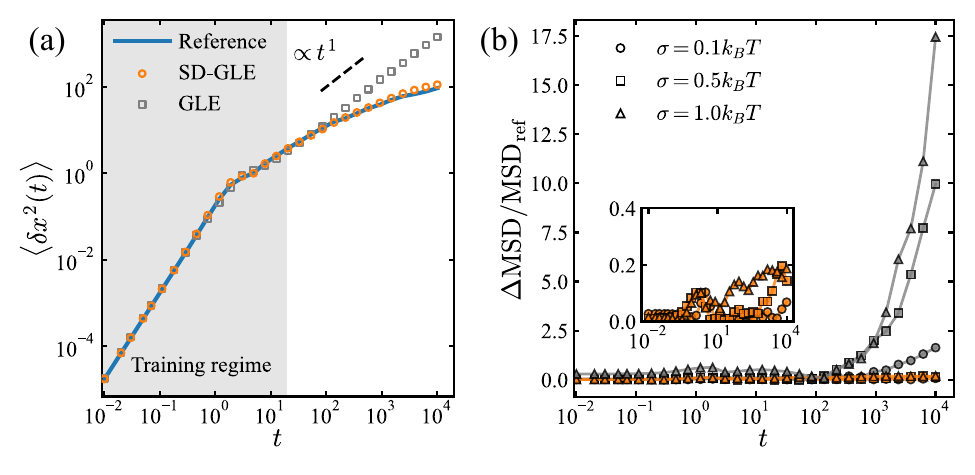}
\caption{(a) Ensemble-averaged MSD generated from simulations using the inferred model parameters compared with the true result. The shaded region indicates the time window of the training data. (b) Relative absolute error of MSD for trajectories with different disorder strengths $\sigma$ in the spatial correlation function $C(r - r')$, where the standard GLE and the SD-GLE are shown in gray and orange respectively.}
\label{fig:2}
\end{figure}

This accurate decoupling is essential for reliably extrapolating macroscopic transport. Fig.~\ref{fig:2}(a) presents the ensemble-averaged mean square displacement (MSD), $\langle \delta x^2(t) \rangle$, extrapolated from a short-time training window (shaded region). While both models capture the transient diffusion, their asymptotic behaviors strictly diverge. Because the standard GLE encodes static trapping into the memory kernel, it assumes that the effects of the rugged landscape vanish over time. Once this artificial memory decays, the system mathematically escapes confinement, forcing an unphysical transition to normal Fickian diffusion and overestimating the long-time mean squared displacement. Conversely, by confining particles within the correct static rugged landscape, the SD-GLE sustains the prolonged anomalous diffusion, tracking the reference MSD for over a temporal range exceeding $10^4$. This physical robustness is quantified by the relative prediction error $ |\text{MSD}_{\text{pred}}(t) - \text{MSD}_{\text{ref}}(t)| / \text{MSD}_{\text{ref}}(t)$ across varying disorder strengths $\sigma$ [Fig.~\ref{fig:2}(b)]. As $\sigma$ increases, the mean-field approximation in the standard GLE becomes inaccurate, leading to a growing error of MSD. In contrast, the SD-GLE maintains a small and stable error across all $\sigma$, indicating that separating spatial disorder from dynamic friction is important for predicting long-time behavior in strongly heterogeneous systems.

\begin{figure}
\centering
\includegraphics[width=1.0\linewidth]{fig1.pdf}
\caption{Ideal 1D case. (a) Memory kernels $K(t)$ inferred from the SD-GLE and GLE model are compared with the true kernel used to generate the trajectory data, (b) Posterior distribution of the effective potential inferred by our method for a representative replica, compared with the neural network potential inferred from the GLE model. The shaded region denotes the $\pm 2$ standard deviation credible interval of the posterior, while the scatter points indicate the mean values of the learned inducing points $\mathbf{u}^{(\alpha)}$. The inset compares the potential correlation function $C(\left|r-r'\right|)$ inferred from the SD-GLE model with the reference.}
\label{fig:1}
\end{figure}

Then we perform molecular dynamics simulations of a two-dimensional Kob-Andersen (KA) binary mixture \cite{2012_KA,2016_KA_dynamics} comprising $N = 1000$ particles with a composition ratio of 65:35. The system is first melted at high temperature before thoroughly equilibrated in the target temperature under the NVT ensemble. A fraction $c=0.04$ of the particles are randomly selected and pinned for each independent replica. Production simulations for the remaining mobile particles are conducted in the NVE ensemble. The trajectories of individual unpinned A-type particles along a single axis $x$ are extracted as the CG coordinates. All following physical quantities are reported
in reduced Lennard–Jones units, with $\sigma$, $\epsilon$, $\sigma (m/\epsilon)^{1/2}$ being
the unit of length, energy, and time.

\begin{figure}
\centering
\includegraphics[width=1.0\linewidth]{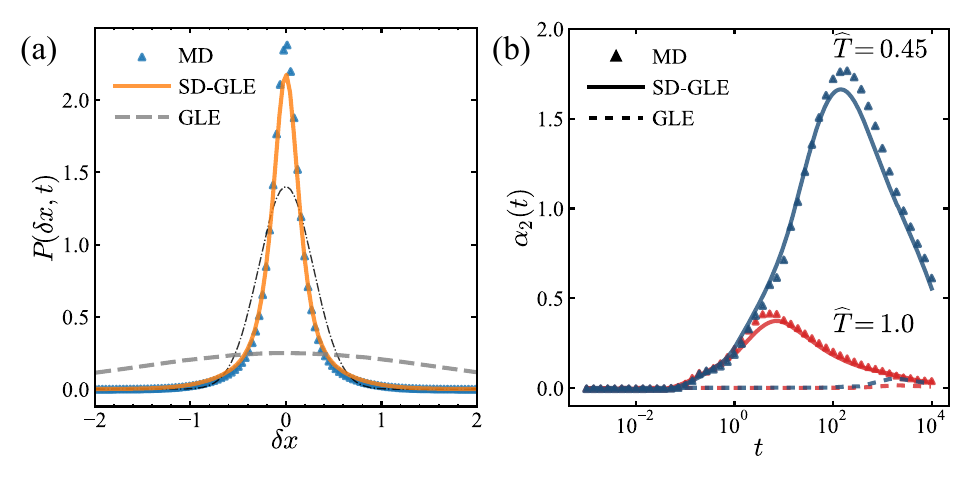}
\caption{(a) Ensemble-averaged probability density functions of the displacements at supercool state ($\widehat{T} = 0.45$) and time $t = 200$. The dash-dotted black line represents a Gaussian distribution with identical variance of the MD data. (b) Time evolution of the non-Gaussian parameter $\alpha_2(t)$ within two characteristic temperatures.}
\label{fig:3}
\end{figure}

To validate the predictive capability of our framework, we perform ensemble simulations driven by the inferred spatial disorder and memory kernel, and first examine the resulting displacement probability density function, $P(\delta x, t)$. In the deeply supercooled state ($\widehat{T}=k_BT/\epsilon=0.45$), the distribution at time $t=200$ exhibits a sharp central peak and broad tails, indicating significant deviations from Gaussian behavior (dash-dotted line in Fig.~\ref{fig:3}(a)). The SD-GLE quantitatively reproduces the MD data, whereas the standard GLE, governed by stationary Gaussian noise, fails to resolve these heterogeneous features. Although the standard GLE's deviations are visually subtle in $P(\delta x, t)$, its failure is quantified by the non-Gaussian parameter \cite{1964_NGP}, $\alpha_2(t) = \frac{\langle \delta x^4 \rangle}{3 \langle \delta x^2 \rangle^2} - 1$. As illustrated in Fig.~\ref{fig:3}(b), the $\alpha_2(t)$ calculated from MD simulations exhibits a non-monotonic evolution, whose peak grows as temperature decreases, reflecting enhanced dynamic heterogeneity. The standard GLE yields $\alpha_2(t) \approx 0$ at all times, demonstrating that a stationary memory kernel alone cannot generate non-Gaussianity. By integrating static spatial disorder, the SD-GLE quantitatively reproduces the peak position and amplitude of $\alpha_2(t)$ at both $\widehat{T}=0.45$ and $\widehat{T}=1.0$. This demonstrates that explicitly mapping the rugged spatial topography is not merely a geometric correction, but the fundamental physical requisite for recovering the underlying dynamic heterogeneity, which ultimately drives the heavy-tailed anomalous transport missed by homogeneous bath models.

\begin{figure}
\centering
\includegraphics[width=1.0\linewidth]{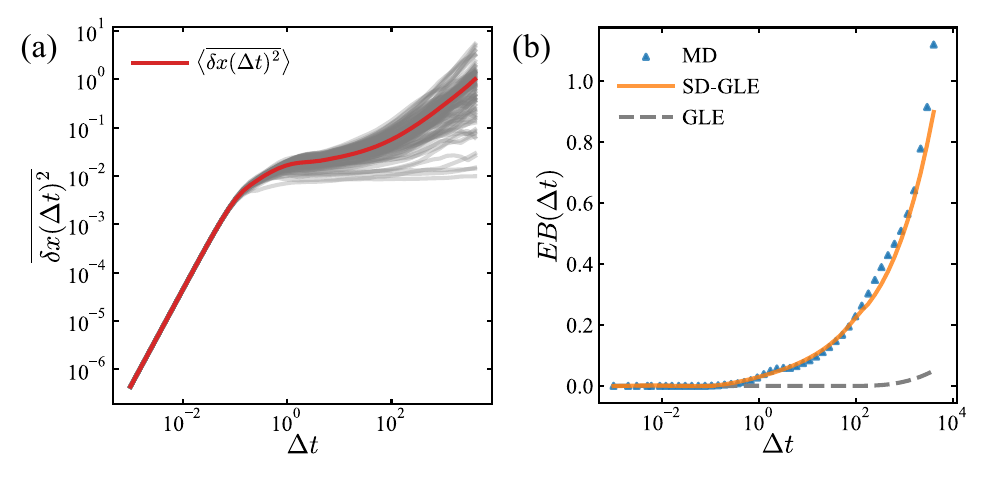}
\caption{(a) Distribution of time-averaged MSDs for individual trajectories at a representative lag time. $\langle \overline{{\delta x}(\Delta t)^2} \rangle$ denotes the ensemble average of time-averaged MSD over 200 trajectories. (b) The ergodicity breaking parameter EB as a function of lag time $\Delta t$.}
\label{fig:4}
\end{figure}

Finally, Fig.~\ref{fig:4}(a) displays the time-averaged MSDs, $\overline{\delta^2(\Delta t)}$, of individual MD trajectories at $\widehat{T}=0.45$ as a function of the lag time $\Delta t$. Instead of converging to the ensemble average, the time-averaged MSDs exhibit significant dispersion at large lag times $\Delta t$. This non-self-averaging behavior indicates weak ergodicity breaking, arising from particles exploring distinct local environments dictated by spatial disorder. To quantify this, we evaluate the ergodicity breaking parameter, $\mathrm{EB}(\Delta t) = \frac{\langle [\overline{\delta^2(\Delta t)}]^2 \rangle}{\langle \overline{\delta^2(\Delta t)} \rangle^2} - 1$, in Fig.~\ref{fig:4}(b) \cite{2008_EB}, where $\langle\cdot\rangle$ refers to ensemble average over all trajectories. The MD data yield a growing $\mathrm{EB}(\Delta t)$, confirming persistent non-ergodic transport. Constrained by a homogeneous bath assumption, the standard GLE predicts purely ergodic behavior ($\mathrm{EB} \approx 0$). By incorporating static landscape fluctuations, the SD-GLE quantitatively reproduces the temporal evolution of $\mathrm{EB}$. The presence of weak ergodicity breaking exposes a limitation of conventional ensemble-based inference, as a single globally shared equation of motion may fail to capture the resulting heterogeneous dynamics.

To conclude, we have introduced the SD-GLE framework, which employs variational inference to quantitatively disentangle static spatial disorder from viscoelastic memory in heterogeneous systems. By parameterizing a coarse-grained model with these extracted features, we accurately capture the crossover in diffusion dynamics from short-time trajectory ensembles. The framework goes beyond mean-field descriptions and captures dynamical heterogeneity as well as fluctuations between individual trajectories. Furthermore, our results indicate that explicitly separating static spatial disorder from dynamic friction is important for constructing predictive data-driven dynamical models. Because of its adaptability and scalability, the SD-GLE provides a robust foundation for predicting the long-term dynamics of complex systems directly from limited observations.

~\\
\textit{Data Availability Statement}---The source code for reproducing the findings in this paper is available at \url{https://github.com/aweqardf/Spatial_disorder_GLE}. 

~\\
\textit{Acknowledgments}---This work was supported by the National Natural Science Foundation of China (Grant No. 12404291 and No. 32371299), National Key Research and Development Program of China (Grant No. 2025YFC2311702), and New Generation Artificial Intelligence-National Science and Technology Major Project (2025ZD0121802).
C.L. did this work during his internship at Shanghai Artificial Intelligence Laboratory. 

\bibliographystyle{apsrev4-2}
\bibliography{refs}

@PREAMBLE{
 "\providecommand{\noopsort}[1]{}" 
 # "\providecommand{\singleletter}[1]{#1}%" 
}

@article{2004_Bio_Anomalous,
  title = {Anomalous Diffusion in Living Yeast Cells},
  author = {Toli\ifmmode \acute{c}\else \'{c}\fi{}-N\o{}rrelykke, Iva Marija and Munteanu, Emilia-Laura and Thon, Genevieve and Oddershede, Lene and Berg-S\o{}rensen, Kirstine},
  journal = {Phys. Rev. Lett.},
  volume = {93},
  issue = {7},
  pages = {078102},
  numpages = {4},
  year = {2004},
  month = {Aug},
  publisher = {American Physical Society},
  doi = {10.1103/PhysRevLett.93.078102},
  url = {https://link.aps.org/doi/10.1103/PhysRevLett.93.078102}
}

@article{
2021_Bio_Anomalous,
author = {Henrik D. Pinholt  and Søren S.-R. Bohr  and Josephine F. Iversen  and Wouter Boomsma  and Nikos S. Hatzakis },
title = {Single-particle diffusional fingerprinting: A machine-learning framework for quantitative analysis of heterogeneous diffusion},
journal = {Proceedings of the National Academy of Sciences},
volume = {118},
number = {31},
pages = {e2104624118},
year = {2021},
doi = {10.1073/pnas.2104624118},
URL = {https://www.pnas.org/doi/abs/10.1073/pnas.2104624118}}

@article{2023_Polymer_Anomalous,
  title = {Data-Driven Learning of the Generalized Langevin Equation with State-Dependent Memory},
  author = {Ge, Pei and Zhang, Zhongqiang and Lei, Huan},
  journal = {Phys. Rev. Lett.},
  volume = {133},
  issue = {7},
  pages = {077301},
  numpages = {6},
  year = {2024},
  month = {Aug},
  publisher = {American Physical Society},
  doi = {10.1103/PhysRevLett.133.077301},
  url = {https://link.aps.org/doi/10.1103/PhysRevLett.133.077301}
}

@article{2024_Polymer_Anomalous,
author = {Baicheng Mei  and Gary S. Grest  and Songyue Liu  and Thomas C. O’Connor  and Kenneth S. Schweizer },
title = {Unified understanding of the impact of semiflexibility, concentration, and molecular weight on macromolecular-scale ring diffusion},
journal = {Proceedings of the National Academy of Sciences},
volume = {121},
number = {31},
pages = {e2403964121},
year = {2024},
doi = {10.1073/pnas.2403964121},
URL = {https://www.pnas.org/doi/abs/10.1073/pnas.2403964121}}

@article{2023_Glass_Anomalous,
  title = {Predicting Dynamic Heterogeneity in Glass-Forming Liquids by Physics-Inspired Machine Learning},
  author = {Jung, Gerhard and Biroli, Giulio and Berthier, Ludovic},
  journal = {Phys. Rev. Lett.},
  volume = {130},
  issue = {23},
  pages = {238202},
  numpages = {6},
  year = {2023},
  month = {Jun},
  publisher = {American Physical Society},
  doi = {10.1103/PhysRevLett.130.238202},
  url = {https://link.aps.org/doi/10.1103/PhysRevLett.130.238202}
}

@article{2020_Glass_Anomalous,
  author  = {Bapst, Victor and Keck, Tilman and Grabska-Barwi{\'n}ska, Agnieszka and Donner, Christian and Cubuk, Ekin D. and Schoenholz, Samuel S. and Obika, Akinori and Nelson, Andrew W. R. and Back, Thomas and Hassabis, Demis and Kohli, Pushmeet},
  title   = {Unveiling the predictive power of static structure in glassy systems},
  journal = {Nature Physics},
  year    = {2020},
  volume  = {16},
  number  = {4},
  pages   = {448--454},
  doi     = {10.1038/s41567-020-0842-8},
  url     = {https://doi.org/10.1038/s41567-020-0842-8},
  issn    = {1745-2481}
}

@article{2022Likelihood_Memory,
author = {Hadrien Vroylandt  and Ludovic Goudenège  and Pierre Monmarché  and Fabio Pietrucci  and Benjamin Rotenberg },
title = {Likelihood-based non-Markovian models from molecular dynamics},
journal = {Proceedings of the National Academy of Sciences},
volume = {119},
number = {13},
pages = {e2117586119},
year = {2022},
doi = {10.1073/pnas.2117586119},
}

@article{Tepper2024MemoryKernel,
  title   = {Accurate Memory Kernel Extraction from Discretized Time-Series Data},
  author  = {Tepper, L. and Dalton, B. and Netz, R. R.},
  journal = {Journal of Chemical Theory and Computation},
  year    = {2024},
  volume  = {20},
  number  = {8},
  pages   = {3061--3068},
  doi     = {10.1021/acs.jctc.3c01289}
}

@article{2011_Two_Mechanisms,
  author  = {Weigel, Adrian V. and Simon, Brian and Tamkun, Michael M. and Krapf, Diego},
  title   = {Ergodic and nonergodic processes coexist in the plasma membrane as observed by single-molecule tracking},
  journal = {Proceedings of the National Academy of Sciences of the United States of America},
  year    = {2011},
  volume  = {108},
  number  = {16},
  pages   = {6438--6443},
  doi     = {10.1073/pnas.1016325108},
  url     = {https://doi.org/10.1073/pnas.1016325108},
  pmid    = {21464280},
  pmcid   = {PMC3081000}
}

@article{2010_Two_Mechanisms,
  title = {Subdiffusion of mixed origins: When ergodicity and nonergodicity coexist},
  author = {Meroz, Yasmine and Sokolov, Igor M. and Klafter, Joseph},
  journal = {Phys. Rev. E},
  volume = {81},
  issue = {1},
  pages = {010101},
  numpages = {4},
  year = {2010},
  month = {Jan},
  publisher = {American Physical Society},
  doi = {10.1103/PhysRevE.81.010101},
  url = {https://link.aps.org/doi/10.1103/PhysRevE.81.010101}
}

@article{2014_Diffusion_Random_Potential,
  title = {Anomalous Features of Diffusion in Corrugated Potentials with Spatial Correlations: Faster than Normal, and Other Surprises},
  author = {Goychuk, Igor and Kharchenko, Vasyl O.},
  journal = {Phys. Rev. Lett.},
  volume = {113},
  issue = {10},
  pages = {100601},
  numpages = {5},
  year = {2014},
  month = {Sep},
  publisher = {American Physical Society},
  doi = {10.1103/PhysRevLett.113.100601},
  url = {https://link.aps.org/doi/10.1103/PhysRevLett.113.100601}
}

@article{2013_Indistinguishable,
  title = {Disentangling Sources of Anomalous Diffusion},
  author = {Thiel, Felix and Flegel, Franziska and Sokolov, Igor M.},
  journal = {Phys. Rev. Lett.},
  volume = {111},
  issue = {1},
  pages = {010601},
  numpages = {4},
  year = {2013},
  month = {Jul},
  publisher = {American Physical Society},
  doi = {10.1103/PhysRevLett.111.010601},
  url = {https://link.aps.org/doi/10.1103/PhysRevLett.111.010601}
}

@article{2021_ML_Distinguish,
  author  = {Mu{\~n}oz-Gil, G. and Volpe, G. and Garcia-March, M. A. and others},
  title   = {Objective comparison of methods to decode anomalous diffusion},
  journal = {Nature Communications},
  year    = {2021},
  volume  = {12},
  pages   = {6253},
  doi     = {10.1038/s41467-021-26320-w},
  url     = {https://doi.org/10.1038/s41467-021-26320-w}
}

@article{2016_Single_trajectory,
  title = {Universal Fluctuations of Single-Particle Diffusivity in a Quenched Environment},
  author = {Akimoto, Takuma and Barkai, Eli and Saito, Keiji},
  journal = {Phys. Rev. Lett.},
  volume = {117},
  issue = {18},
  pages = {180602},
  numpages = {5},
  year = {2016},
  month = {Oct},
  publisher = {American Physical Society},
  doi = {10.1103/PhysRevLett.117.180602},
  url = {https://link.aps.org/doi/10.1103/PhysRevLett.117.180602}
}

@article{2008_Ergodicity_Breaking,
  title = {Nonergodicity Mimics Inhomogeneity in Single Particle Tracking},
  author = {Lubelski, Ariel and Sokolov, Igor M. and Klafter, Joseph},
  journal = {Phys. Rev. Lett.},
  volume = {100},
  issue = {25},
  pages = {250602},
  numpages = {4},
  year = {2008},
  month = {Jun},
  publisher = {American Physical Society},
  doi = {10.1103/PhysRevLett.100.250602},
  url = {https://link.aps.org/doi/10.1103/PhysRevLett.100.250602}
}

@article{2011_Ergodicity_Breaking,
  title = {In Vivo Anomalous Diffusion and Weak Ergodicity Breaking of Lipid Granules},
  author = {Jeon, Jae-Hyung and Tejedor, Vincent and Burov, Stas and Barkai, Eli and Selhuber-Unkel, Christine and Berg-S\o{}rensen, Kirstine and Oddershede, Lene and Metzler, Ralf},
  journal = {Phys. Rev. Lett.},
  volume = {106},
  issue = {4},
  pages = {048103},
  numpages = {4},
  year = {2011},
  month = {Jan},
  publisher = {American Physical Society},
  doi = {10.1103/PhysRevLett.106.048103},
  url = {https://link.aps.org/doi/10.1103/PhysRevLett.106.048103}
}

@Article{2010_MZ_for_MD,
author ="Hijón, Carmen and Español, Pep and Vanden-Eijnden, Eric and Delgado-Buscalioni, Rafael",
title  ="Mori–Zwanzig formalism as a practical computational tool",
journal  ="Faraday Discuss.",
year  ="2010",
volume  ="144",
issue  ="0",
pages  ="301-322",
publisher  ="The Royal Society of Chemistry",
doi  ="10.1039/B902479B",
url  ="http://dx.doi.org/10.1039/B902479B"}

@article{2017_Memory,
  title = {External Potential Modifies Friction of Molecular Solutes in Water},
  author = {Daldrop, Jan O. and Kowalik, Bartosz G. and Netz, Roland R.},
  journal = {Phys. Rev. X},
  volume = {7},
  issue = {4},
  pages = {041065},
  numpages = {14},
  year = {2017},
  month = {Dec},
  publisher = {American Physical Society},
  doi = {10.1103/PhysRevX.7.041065},
  url = {https://link.aps.org/doi/10.1103/PhysRevX.7.041065}
}

@article{2023_HuanLei_Memory,
  title = {Construction of Coarse-Grained Molecular Dynamics with Many-Body Non-Markovian Memory},
  author = {Lyu, Liyao and Lei, Huan},
  journal = {Phys. Rev. Lett.},
  volume = {131},
  issue = {17},
  pages = {177301},
  numpages = {6},
  year = {2023},
  month = {Oct},
  publisher = {American Physical Society},
  doi = {10.1103/PhysRevLett.131.177301},
  url = {https://link.aps.org/doi/10.1103/PhysRevLett.131.177301}
}

@article{2024_AIGLE,
author = {Pinchen Xie  and Roberto Car  and Weinan E },
title = {Ab initio generalized Langevin equation},
journal = {Proceedings of the National Academy of Sciences},
volume = {121},
number = {14},
pages = {e2308668121},
year = {2024},
doi = {10.1073/pnas.2308668121},
URL = {https://www.pnas.org/doi/abs/10.1073/pnas.2308668121}}

@InProceedings{2009_SVGP,
  title = 	 {Variational Learning of Inducing Variables in Sparse Gaussian Processes},
  author = 	 {Titsias, Michalis},
  booktitle = 	 {Proceedings of the Twelfth International Conference on Artificial Intelligence and Statistics},
  pages = 	 {567--574},
  year = 	 {2009},
  editor = 	 {van Dyk, David and Welling, Max},
  volume = 	 {5},
  month = 	 {16--18 Apr},
  publisher =    {PMLR},
  url = 	 {https://proceedings.mlr.press/v5/titsias09a.html}
}

@article{2016_KA_dynamics,
    author = {Chakrabarty, Saurish and Das, Rajsekhar and Karmakar, Smarajit and Dasgupta, Chandan},
    title = {Understanding the dynamics of glass-forming liquids with random pinning within the random first order transition theory},
    journal = {The Journal of Chemical Physics},
    volume = {145},
    number = {3},
    pages = {034507},
    year = {2016},
    month = {07},
    issn = {0021-9606},
    doi = {10.1063/1.4958632},
    url = {https://doi.org/10.1063/1.4958632},
}

@article{
2012_KA,
author = {Chiara Cammarota  and Giulio Biroli },
title = {Ideal glass transitions by random pinning},
journal = {Proceedings of the National Academy of Sciences},
volume = {109},
number = {23},
pages = {8850-8855},
year = {2012},
doi = {10.1073/pnas.1111582109},
URL = {https://www.pnas.org/doi/abs/10.1073/pnas.1111582109}}

@article{1964_NGP,
  title = {Correlations in the Motion of Atoms in Liquid Argon},
  author = {Rahman, A.},
  journal = {Phys. Rev.},
  volume = {136},
  issue = {2A},
  pages = {A405--A411},
  numpages = {0},
  year = {1964},
  month = {Oct},
  publisher = {American Physical Society},
  doi = {10.1103/PhysRev.136.A405},
  url = {https://link.aps.org/doi/10.1103/PhysRev.136.A405}
}

@article{2008_EB,
  title = {Random Time-Scale Invariant Diffusion and Transport Coefficients},
  author = {He, Y. and Burov, S. and Metzler, R. and Barkai, E.},
  journal = {Phys. Rev. Lett.},
  volume = {101},
  issue = {5},
  pages = {058101},
  numpages = {4},
  year = {2008},
  month = {Jul},
  publisher = {American Physical Society},
  doi = {10.1103/PhysRevLett.101.058101},
  url = {https://link.aps.org/doi/10.1103/PhysRevLett.101.058101}
}

@article{2010_Markov_embedding,
author = {Ceriotti, Michele and Bussi, Giovanni and Parrinello, Michele},
title = {Colored-Noise Thermostats à la Carte},
journal = {Journal of Chemical Theory and Computation},
volume = {6},
number = {4},
pages = {1170-1180},
year = {2010},
doi = {10.1021/ct900563s},
URL = {    
        https://doi.org/10.1021/ct900563s
},
}

@article{1990_CLT,
title = {Anomalous diffusion in disordered media: Statistical mechanisms, models and physical applications},
journal = {Physics Reports},
volume = {195},
number = {4},
pages = {127-293},
year = {1990},
issn = {0370-1573},
doi = {https://doi.org/10.1016/0370-1573(90)90099-N},
url = {https://www.sciencedirect.com/science/article/pii/037015739090099N},
author = {Jean-Philippe Bouchaud and Antoine Georges}
}

@article{
2016_Data_Driven_GLE,
author = {Huan Lei  and Nathan A. Baker  and Xiantao Li },
title = {Data-driven parameterization of the generalized Langevin equation},
journal = {Proceedings of the National Academy of Sciences},
volume = {113},
number = {50},
pages = {14183-14188},
year = {2016},
doi = {10.1073/pnas.1609587113},
URL = {https://www.pnas.org/doi/abs/10.1073/pnas.1609587113}}

@ARTICLE{1965_KF,
author={Schweppe, F.},
journal={IEEE Transactions on Information Theory}, 
title={Evaluation of likelihood functions for Gaussian signals}, 
year={1965},
volume={11},
number={1},
pages={61-70},
doi={10.1109/TIT.1965.1053737}}

@article{2010_Markov_Embedding_Temp,
  title = {Langevin Equation with Colored Noise for Constant-Temperature Molecular Dynamics Simulations},
  author = {Ceriotti, Michele and Bussi, Giovanni and Parrinello, Michele},
  journal = {Phys. Rev. Lett.},
  volume = {102},
  issue = {2},
  pages = {020601},
  numpages = {4},
  year = {2009},
  month = {Jan},
  publisher = {American Physical Society},
  doi = {10.1103/PhysRevLett.102.020601},
  url = {https://link.aps.org/doi/10.1103/PhysRevLett.102.020601}
}

@article{1987_Supercool_GRF,
title = {Viscous flow in supercooled liquids analyzed in terms of transport theory for random media with energetic disorder},
author = {B\"assler, H.},
journal = {Phys. Rev. Lett.},
volume = {58},
issue = {8},
pages = {767--770},
numpages = {0},
year = {1987},
month = {Feb},
publisher = {American Physical Society},
doi = {10.1103/PhysRevLett.58.767},
url = {https://link.aps.org/doi/10.1103/PhysRevLett.58.767}
}

@article{2008_Supercool_GRF,
author  = {Hecksher, Tina and Nielsen, Albena I. and Olsen, Niels Boye and Dyre, Jeppe C.},
title   = {Little evidence for dynamic divergences in ultraviscous molecular liquids},
journal = {Nature Physics},
year    = {2008},
volume  = {4},
number  = {9},
pages   = {737--741},
doi     = {10.1038/nphys1033},
url     = {https://doi.org/10.1038/nphys1033},
issn    = {1745-2481}
}

@article{2016_Protein_GRF,
author  = {Kong, Muwen and Liu, Lili and Chen, Xuejing and Driscoll, Katherine I. and Mao, Peng and B{\"o}hm, Stefanie and Kad, Neil M. and Watkins, Simon C. and Bernstein, Kara A. and Wyrick, John J. and Min, Jung-Hyun and Van Houten, Bennett},
title   = {Single-Molecule Imaging Reveals that Rad4 Employs a Dynamic DNA Damage Recognition Process},
journal = {Molecular Cell},
year    = {2016},
volume  = {64},
number  = {2},
pages   = {376--387},
doi     = {10.1016/j.molcel.2016.09.005},
url     = {https://doi.org/10.1016/j.molcel.2016.09.005},
issn    = {1097-2765},
publisher = {Elsevier}
}

@article{
2002_Protein_GRF,
author = {Ulrich Gerland  and J. David Moroz  and Terence Hwa },
title = {Physical constraints and functional characteristics of transcription factor–DNA interaction},
journal = {Proceedings of the National Academy of Sciences},
volume = {99},
number = {19},
pages = {12015-12020},
year = {2002},
doi = {10.1073/pnas.192693599},
URL = {https://www.pnas.org/doi/abs/10.1073/pnas.192693599}}

@article{2004_Protein_GRF,
  title = {Diffusion in correlated random potentials, with applications to DNA},
  author = {Slutsky, Michael and Kardar, Mehran and Mirny, Leonid A.},
  journal = {Phys. Rev. E},
  volume = {69},
  issue = {6},
  pages = {061903},
  numpages = {11},
  year = {2004},
  month = {Jun},
  publisher = {American Physical Society},
  doi = {10.1103/PhysRevE.69.061903},
  url = {https://link.aps.org/doi/10.1103/PhysRevE.69.061903}
}

@article{2009_Protein_GRF,
  title = {Searching Fast for a Target on DNA without Falling to Traps},
  author = {B\'enichou, O. and Kafri, Y. and Sheinman, M. and Voituriez, R.},
  journal = {Phys. Rev. Lett.},
  volume = {103},
  issue = {13},
  pages = {138102},
  numpages = {4},
  year = {2009},
  month = {Sep},
  publisher = {American Physical Society},
  doi = {10.1103/PhysRevLett.103.138102},
  url = {https://link.aps.org/doi/10.1103/PhysRevLett.103.138102}
}

@article{2012_Protein_GRF,
doi = {10.1088/0034-4885/75/2/026601},
url = {https://doi.org/10.1088/0034-4885/75/2/026601},
year = {2012},
month = {jan},
publisher = {},
volume = {75},
number = {2},
pages = {026601},
author = {Sheinman, M and Bénichou, O and Kafri, Y and Voituriez, R},
title = {Classes of fast and specific search mechanisms for proteins on DNA},
journal = {Reports on Progress in Physics}
}

@article{2025_GLE_with_Potential,
  author  = {Klimek, Anton and Dalton, Benjamin A. and Netz, Roland R.},
  title   = {Subdiffusion from competition between multi-exponential friction memory and energy barriers},
  journal = {The European Physical Journal E},
  year    = {2025},
  volume  = {48},
  number  = {8},
  pages   = {55},
  doi     = {10.1140/epje/s10189-025-00518-y},
  url     = {https://doi.org/10.1140/epje/s10189-025-00518-y},
  issn    = {1292-895X}
}

@article{2023_Distinguish,
  title = {Compression algorithms reveal memory effects and static disorder in single-molecule trajectories},
  author = {Song, Kevin and Makarov, Dmitrii E. and Vouga, Etienne},
  journal = {Phys. Rev. Res.},
  volume = {5},
  issue = {1},
  pages = {L012026},
  numpages = {7},
  year = {2023},
  month = {Feb},
  publisher = {American Physical Society},
  doi = {10.1103/PhysRevResearch.5.L012026},
  url = {https://link.aps.org/doi/10.1103/PhysRevResearch.5.L012026}
}

@article{2015_two_mechanisms_toolbox,
  title={A toolbox for determining subdiffusive mechanisms},
  author={Meroz, Yael and Sokolov, Igor M},
  journal={Physics Reports},
  volume={573},
  pages={1--29},
  year={2015},
  publisher={Elsevier}
}

@article{2014_anomalous_diffusion,
  title={Anomalous diffusion models and their properties: non-stationarity, non-ergodicity, and ageing at the centenary of single particle tracking},
  author={Metzler, Ralf and Jeon, Jae-Hyung and Cherstvy, Andrey G and Barkai, Eli},
  journal={Physical Chemistry Chemical Physics},
  volume={16},
  number={44},
  pages={24128--24164},
  year={2014},
  publisher={Royal Society of Chemistry}
}

@article{2021_diffusion_objective,
  title={Objective comparison of methods to decode anomalous diffusion},
  author={Mu{\~n}oz-Gil, Gorka and Volpe, Giovanni and Garcia-March, Miguel Angel and Aghion, Erez and Argun, Aykut and Hong, Changbiao and Bland, Tom and Bo, Stefano and Conejero, J Alberto and Firbas, N and others},
  journal={Nature communications},
  volume={12},
  number={1},
  pages={6253},
  year={2021},
  publisher={Nature Publishing Group UK London}
}

@article{2019_classification,
  title={Classification of diffusion modes in single-particle tracking data: Feature-based versus deep-learning approach},
  author={Kowalek, Piotr and Loch-Olszewska, Hanna and Szwabi{\'n}ski, Janusz},
  journal={Physical Review E},
  volume={100},
  number={3},
  pages={032410},
  year={2019},
  publisher={APS}
}

@article{2019_measurement,
  title={Measurement of anomalous diffusion using recurrent neural networks},
  author={Bo, Stefano and Schmidt, Falko and Eichhorn, Ralf and Volpe, Giovanni},
  journal={Physical Review E},
  volume={100},
  number={1},
  pages={010102},
  year={2019},
  publisher={APS}
}

@article{2024_coarse_grain,
  title={Coarse-graining conformational dynamics with multidimensional generalized Langevin equation: How, when, and why},
  author={Jiang, Yi and E, Weinan},
  journal={Journal of Chemical Theory and Computation},
  volume={20},
  number={7},
  pages={2824--2836},
  year={2024},
  publisher={ACS Publications}
}

@article{2012_polymer,
  title={When Brownian diffusion is not Gaussian},
  author={Wang, Bo and Kuo, Stephen M and Bae, Sung Chul and Granick, Steve},
  journal={Nature materials},
  volume={11},
  number={6},
  pages={481--485},
  year={2012},
  publisher={Nature Publishing Group}
}

@Article{2012_living_cell,
author ="Sokolov, Igor M.",
title  ="Models of anomalous diffusion in crowded environments",
journal  ="Soft Matter",
year  ="2012",
volume  ="8",
issue  ="35",
pages  ="9043-9052",
publisher  ="The Royal Society of Chemistry",
doi  ="10.1039/C2SM25701G",
url  ="http://dx.doi.org/10.1039/C2SM25701G"}

@article{2022_position_dependent,
  title={Position-dependent memory kernel in generalized Langevin equations: theory and numerical estimation},
  author={Gouden{\`e}ge, Ludovic and Leli{\`e}vre, Tony and Pavliotis, Grigorios A and Stoltz, Gabriel},
  journal={Multiscale Modeling \& Simulation},
  volume={20},
  number={4},
  pages={1454--1486},
  year={2022},
  publisher={SIAM}
}

@article{2021_data_driven_GLE,
  title={Model reduction with memory and the machine learning of dynamical systems},
  author={Ma, Chao and Wang, Jianchun and E, Weinan},
  journal={Communications in Computational Physics},
  volume={29},
  number={4},
  pages={1002--1035},
  year={2021},
  publisher={Global Science Press}
}

\end{document}